\documentclass[floats,prd,showpacs,amssymb]{revtex4}
\usepackage{amssymb}
\usepackage{graphics}
\usepackage{amsmath}
\usepackage{amsfonts}
\usepackage{bm}

\begin{document}

\title{Constraints on massive gravity theory from big bang nucleosynthesis}%

\author{G. Lambiase$^{a,b}$} 
\affiliation{$^a$University of Salerno, 84084 - Fisciano (SA), Italy.\\
$^b$INFN, Sezione di Napoli, Italy.}
\def\be{\begin{equation}}
\def\ee{\end{equation}}
\def\al{\alpha}
\def\bea{\begin{eqnarray}}
\def\eea{\end{eqnarray}}

\begin{abstract}
The massive gravity cosmology is studied in the scenario of big bang nucleosynthesis.
By making use of current bounds on the deviation from the fractional mass, we derive the constraints on the free parameters
of the theory. The cosmological consequences of the model are also discussed in the framework of the PAMELA experiment.
\end{abstract}

\maketitle


\section{Introduction}

Recently, de Rham, Gabadadze and Tolley (dRGT) \cite{derham} have proposed a model of gravity in which the mass of the graviton
is taken into account. First attempts in this direction were proposed long time ago by Fierz and Pauli \cite{pauli},
who constructed a (ghost-free) linear theory of massive gravity. As later realized, the Fierz-Pauli theory is affected by some pathologies
as, for example, it is in conflict with solar system tests \cite{solartest}. A renew interest for this theory is arisen in the last years
thanks to the Stueckelberg formalism introduced in \cite{siegel}  (for a review, see \cite{kurt}).

The dRGT model relies on the idea to add higher order self-interaction graviton terms to the Einstein-Hilbert action in order to
get rid of the Boudware-Deser instability \cite{deser}.
An interesting consequence of the theory is that the presence of a tiny mass
of the graviton gives rise to a term that play the role of cosmological constant.
Therefore, as a {\it modified} theory of gravity, it allows to account
for the observed acceleration of the present Universe, without invoking exotic matter.
In this respect, many efforts have been devoted to search for cosmological solutions of the field equations that describe
the expansion of the Universe \cite{derham,volkov,langlois,gong,strauss,kobayashi,gratia,comelli,damico,lin,koyama}.
For a Friedman-Robertson-Walker Universe
 \begin{equation}\label{luineelement}
    ds^2=a^2(t)\left[dt^2-\frac{dr^2}{1-Kr^2}-r^2(d\theta^2+\sin^2\theta d\phi^2)\right]\,.
 \end{equation}
where $a(t)$ is the scale factor and $K=\pm 1, 0$ is the spatial curvature, the modified Friedman equation (the $0-0$ field equation) reads
 \begin{equation}\label{1}
    3\, \frac{{\dot a}^2}{a^4}+\frac{3K}{a^2}=\kappa^2\rho+m^2\left(\alpha_0+\frac{\alpha_1}{a}+\frac{\alpha_2}{a^2}+\frac{\alpha_3}{a^3}\right)\,,
 \end{equation}
where the dot indicates the time derivative, ${\kappa}^2=8\pi/m_P^2$, $m_P=G^{-1/2}=1.22\times 10^{19}$GeV is the Planck mass (in  natural units), and $\alpha_i$, $i=0,1,2,3$ are constants
 \begin{equation}\label{constantsalpha}
    \alpha_0 \equiv c_4-4c_3-6\,,\quad  \alpha_1\equiv 3C(3+3c_3-c_4)\,, \quad \alpha_2 \equiv 3C^2(c_4-2c_3-1)\,, \quad \alpha_3\equiv C^3(c_3-c_4)\,,
 \end{equation}
where $c_3$ and $c_4$ are the parameters of the model, and $C$ is a constant of integration. The $i-j$ cosmological equations are not relevant for our aim,
and therefore will be not reported here.
The energy-momentum tensor of matter is chosen to be $\kappa^2 T^{\mu}_{\,\,\nu}=diag(\rho, -p, -p, -p)$ and satisfies the equation of continuity
 \begin{equation}\label{nonconserv}
    {\dot \rho}+3H (\rho +p)=0\,,
 \end{equation}
where $H={\dot a}/a$ is the expansion rate of the Universe. Cosmological constraints on massive gravity theory have been recently studied in \cite{cardone}.

The aim of this paper is to study the massive gravity theory in relation to
the formation of the light elements in the early Universe (big bang nucleosynthesis (BBN)).
As well know, the big bang nucleosynthesis provides very stringent constraints on cosmological models, which must be satisfied in order to avoid conflicts with observations.
By making use of the current observational data on primordial
abundance of light elements, we infer constraints on the parameters characterizing the theory.
As an application, we explore the consequences
of the model in the framework of the thermal relic abundance. This analysis is particularly interesting  because alternative cosmologies may modified, in
principle, the thermal history of relic particles. This occurs during the {\it pre} big bang nucleosynthesis epoch, a period of the Universe evolution not directly constrained by cosmological observations.
If the expansion rate of the Universe is enhanced as compared to the expansion rate of general relativity, then thermal relics decouple with larger relic abundance. The change in the Hubble rate may have therefore its imprint on the relic abundance of dark matter, such as WIMPs, axions, heavy neutrinos. These studies have been motivated by the rising behavior of the positron fraction recently observed in the PAMELA experiment \cite{PAMELA6}.

The paper is organized as follows. In Section II we derive the constraints provided by BBN that
the parameters of the massive gravity theory must satisfy in order to be consistent with observational data.
Section III is devoted to the analysis of dark matter relic abundance in the massive gravity cosmology.
Conclusions are shortly discussed in Section IV.

\section{Constraints from Big bang nucleosynthesis}

BBN provides, together with cosmic background radiation,
a strong evidence that the early Universe was hot and dense. In the BBN epoch,
the main interactions of particles are $\nu_e+n  \leftrightarrow  p+e^-$,  $e^++n  \leftrightarrow  p + {\bar \nu}_e$,
$n \leftrightarrow  p+e^- + {\bar \nu}_e$.
The weak interaction rate of particles in thermal equilibrium is given by \cite{bernstein,kolb}
 \begin{equation}\label{Lambdafin}
    \Lambda(T)\simeq q T^5+{\cal O}\left(\frac{\cal Q}{T}\right)\,,
 \end{equation}
where $\Lambda=\Lambda_{\nu_e+n  \leftrightarrow  p+e^-}+\Lambda_{e^++n  \leftrightarrow  p + {\bar \nu}_e}+
\Lambda_{n \leftrightarrow  p+e^- + {\bar \nu}_e}$ and $q=9.6 \times 10^{-46}\text{eV}^{-4}$. Eq. (\ref{Lambdafin}) is obtained in the regime
$T\gg {\cal Q}$, where ${\cal Q}=m_n-m_p$, and $m_{n,p}$ are the neutron and proton masses.
To estimate the primordial mass fraction of ${}^4 He$, one defines the quantity \cite{bernstein,kolb}
 \begin{equation}\label{Yp}
    Y_p\equiv \lambda \, \frac{2 x(t_f)}{1+x(t_f)}\,,
 \end{equation}
where $\lambda=e^{-(t_n-t_f)/\tau}$. $t_f$ and $t_n$ are the time of the freeze-out of the weak interactions and of the nuclesynthesis,
respectively, $\tau\simeq 887$sec is the neutron mean life, and $x(t_f)=e^{-{\cal Q}/T(t_f)}$ is the neutron to proton equilibrium ratio.
The function $\lambda$ represents the fraction of neutrons that decay into protons in the time $t\in [t_f, t_n]$.
Deviations from $Y_p$ (generated by the variation of the freezing temperature $T_f$) are given by
 \begin{equation}\label{deltaYp}
    \delta Y_p=Y_p\left[\left(1-\frac{Y_p}{2\lambda}\right)\ln\left(\frac{2\lambda}{Y_p}-1\right)-\frac{2t_f}{\tau}\right]
    \frac{\delta T_f}{T_f}\,.
 \end{equation}
In the above equation we have set $\delta T(t_n)=0$ because $T_n$ is fixed by the deuterium binding energy \cite{torres}.
By making use of the current estimation on mass fraction $Y_p$ of baryon converted to ${}^4 He$ during the big bang nucleosynthesis \cite{coc,altriBBN}
 \begin{equation}\label{Ypvalues}
 Y_p=0.2476\,, \qquad |\delta Y_p| < 10^{-4}\,,
 \end{equation}
one gets that the upper bound on $\frac{\delta T_f}{T_f}$ is given by
 \begin{equation}\label{deltaT/Tbound}
    \left|\frac{\delta T_f}{T_f}\right| < 4.7 \times 10^{-4}\,.
 \end{equation}
Some comments are in order. Typically during the big bang nucleosynthesis the usual curvature term $K/a^2$ was unimportant, so that we shall set $K=0$.
Moreover, the Universe is radiation dominated with an energy density that scales with $a^4$, i.e. $\rho_r\sim a^{-4}$, or
in terms of the temperature, $\rho_r =\frac{\pi^2 g_*}{30} T^4$.
The energy density $\rho_r$ is conveniently written in terms of the energy ratio $\Omega_r =
\frac{\rho_r}{\rho_{cr}}= \Omega_r^{(0)}(1+z)^4$, where $\Omega_r^{(0)}\sim 10^{-5}$ is the present value of $\Omega_r$,
$\rho_{cr}\simeq 8.4 h^2 10^{-47}$ GeV$^4$
is the critical density ($0.5 < h < 0.8$ is the normalized Hubble constant).

Our aim now is to relate the variation $\delta T_f$ to the parameters (\ref{constantsalpha}) of the theory.
We discuss the cases in which $C\neq 0$ and $C=0$.

\subsection{The case $C\neq 0$}

As shown in Eq. (\ref{1}), the $m^2$-corrections scales as $a^p$, with $p=0, 1, 2, 3$.
In the early Universe, the dominant term is given by $a^3$-term. In the next, the latter
will be treated a perturbation to the radiation energy density $\rho_r$.
Accordingly, the Hubble expansion rate $H\equiv {\dot a}/a$ given by Eq. (\ref{1}) can be written as
 \begin{equation}\label{Hlambda}
    H=\sqrt{1+\frac{m^2\alpha_3}{\kappa^2 \rho a^3}}H_{GR}=H_{GR}+\left(\sqrt{1+\frac{m^2\alpha_3}{\kappa^2 \rho a^3}}-1\right)H_{GR}\,,
 \end{equation}
where $H_{GR}=(\kappa^2 \rho a^2/3)^{1/2}$ is the expansion rate of general relativity.

The freeze-out temperature $T=T_f(1+\frac{\delta T_f}{T_f})$ follows by equating the Hubble expansion rate with
weak interaction rate, $H=\Lambda$. One obtains \cite{kolb}
 \begin{equation}\label{Tf}
    4\times 4! A T_f^5 = H_{GR}(T_f)\quad \to \quad T_f \simeq 0.6\text{MeV}\,,
 \end{equation}
and
\begin{eqnarray}\label{deltaTf}
 \frac{\delta T_f}{T_f} &=& \frac{1}{5}\left(\sqrt{1+\frac{m^2\alpha_3}{\kappa^2 \rho a^3}}-1\right)\simeq 
  \frac{m^2\alpha_3}{10\kappa^2 \rho_{cr}\Omega_r^{(0)}(1+z)} \\
    &\simeq & 1.1 \,\alpha_3  10^{-6}\left(\frac{m}{10^{-42}\text{GeV}}\right)^2
    \left(\frac{0.8}{h}\right)^2\frac{8.4h^210^{-47}\text{GeV}^4}{\rho_{cr}}\,, \label{deltaTf2}
\end{eqnarray}
where we used the upper bound on the graviton mass $m \lesssim 10^{-42}$GeV and $z\simeq 10^9$. As expected, $\delta T_f=0$ in the case $m=0$.

Using the upper bound (\ref{deltaT/Tbound}) one gets
 \begin{equation}\label{alpha3massbound}
    \alpha_3 m^2 < 4.2 \times 10^{-82}\, \text{GeV}^2\,,
 \end{equation}
or
 \begin{equation}\label{alpha3bound}
    \alpha_3 < 4.2 \times 10^{2}\left(\frac{10^{-42} \text{GeV}}{m}\frac{h}{0.8}\right)^2\frac{\rho_{cr}}{8.4 h^2 10^{-47}\text{GeV}^4}\,.
 \end{equation}

\subsection{The case $C=0$}

In this particular case, it follows that the contribution coming from massive gravity, i.e. $m^2\alpha_0$ (see Eq. (\ref{1})), plays the role
of cosmological constant. As discussed in \cite{volkov}, these terms may give rise to the late time acceleration of the Universe.

The analysis goes along the previous one, with the replacement $1+z\to (1+z)^4$ in (\ref{deltaTf}). Therefore one obtains
 \begin{eqnarray}\label{Halpha0}
    H &=&\sqrt{1+\frac{m^2\alpha_0}{\kappa^2 \rho}}H_{GR}\simeq \left[1+\frac{m^2\alpha_0}{2\kappa^2 \rho} \right]H_{GR}\,, \\
    \frac{\delta T_f}{T_f} &=& \frac{m^2\alpha_0}{10\kappa^2 \rho_{cr}\Omega_r^{(0)}(1+z)^4}\,,
 \end{eqnarray}
from which, using again (\ref{deltaT/Tbound}),
 \begin{equation}\label{alpha0bound}
    \alpha_0 m^2 < 4.2 \times 10^{-55}\,\text{GeV}^2\,.
 \end{equation}

\vspace{0.1in}

Equations (\ref{alpha3massbound}) and (\ref{alpha0bound}) represent the main results of the paper:
they give the upper bound on the parameters characterizing the massive gravity theory that must be satisfied
to have compatibility of the theory, and its predictions, with the present constraints dictated by the observational cosmology.
Notice that these equations hold whatever is the mass of graviton. Given the latter, one can definitively fix the
constants $\alpha_0$ and $\alpha_3$.
In what follows we shall refer to them for studying the PAMELA puzzles.

\section{Conclusions}

Massive gravity theory is an alternative covariant formulation of General Relativity. It relies on the fact that the graviton
is massive and that negative energy states (ghosts) are absent. Interestingly, for an homogenous and isotropic Universe, the
field equations turn out to be modified by a constant term that mimics the cosmological constant and by other terms that scale
with the scale factor, mimicking dust, quintessence and stiff matter.

Starting from the modified Friedman equation (\ref{1}) and using  the recent bounds on the primordial light elements, we have derived
the following upper bounds
$\alpha_3 m^2 < 10^{-82}$GeV$^2$ and $\alpha_0 m^2 < 10^{-55}$GeV$^2$. The former is obtained assuming that in the early
Universe the $m^2/a^3$-contribution is dominant with respect to the other corrections, while the letter is obtained setting $C=0$ (or assuming that
the cosmological constant like term $\sim \alpha_0 m^2$ is dominant).

As final comment, we wish to discuss the interesting problem related to the recent results of the PAMELA experiment, 
i.e. the excess of positron events, which could represent a possible signal for dark matter through dark matter annihilation in our Galaxy.
Theoretical results indicate that the PAMELA data can be understood if the annihilation cross sections
are larger than those obtained in standard cosmology. A possible mechanism comes from alternative cosmologies, which give rise to an enhancement
of the expansion rate of the Universe, hence large annihilation cross sections, being at the same time also compatible with other observations.
To account for the enhancement of the expansion rate, it is usual to write \cite{fornengo,BD1,BD}
$H(T)=A(T) H_{GR}(T)$, where $A(T)$ is the enhancement function and $H_{GR}(T)=\sqrt{\frac{8\pi}{3 m_P^2}\, \rho_r}$, with
$\rho_r = \frac{\pi^2 g_*}{30} T^4$, the expansion rate of the Universe compute in the standard General Relativity. Moreover,
conflicts with big bang nuclesynthesis predictions are avoided by working at temperatures of the Universe greater than the
temperature at which the Hubble rate reenters the standard rate of general relativity
($\simeq 1$MeV). In this regime, $A(T)$ is conveniently parameterized as \cite{fornengo}
 \begin{equation}\label{A(T)T>Tre}
    A(T)=1+\eta\left(\frac{T}{T_F}\right)^\nu\,, 
 \end{equation}
where $T_F$ is a reference temperature at which the WIMPs dark matter freezes out in the standard cosmology ($T_F \simeq 17.3$GeV \cite{fornengo}).
In general, $T_F$ depends on the dark matter mass $m_\chi$. $\eta$ and $\nu$ are free parameters and characterize a specific cosmological model.
The values of the parameter $\eta$ required to explain the PAMELA data are $1 \lesssim \eta \lesssim 10^3$ to which corresponds 
the WIMPs dark matter masses $10^2 \text{GeV}\lesssim m_\chi \lesssim 10^3\text{GeV}$. 
For dark matter masses of the order of $10^2$GeV, the parameter $\eta$ can be also close to zero, i.e.
$m_\chi\sim 10^2\text{GeV} \to   0 \lesssim \eta \ll 1$.
Assuming that the dominant term induced by massive gravity is $\alpha_3 /a^3$, and treating it as a perturbation of $\rho_r$, one gets
Eq. (\ref{A(T)T>Tre}) with $\eta = \alpha_3 \frac{30}{16\pi^3g_*}\frac{m^2 m_P^2}{T_0^3 T_F}$ and $\nu=-1$ 
($ T_0=2.7\,\text{K}=2.4\times 10^{-13}$GeV).
Setting $\eta = 10^{-x}\ll 1$ it follows $\alpha_3 m^2 \ll 6\times  10^{-72-x}$GeV$^2$,
that is, comparing with the upper bound provided by big bang nucleosynthesis (\ref{alpha3massbound}), one needs $x\simeq 9-10$, or
$\eta \simeq 10^{-(9\div 10)}$, in order that the massive gravity theory may explain the PAMELA experiment.
The corresponding WIMPs mass is of the order $10^2$GeV.
In the case in which $C=0$, the amplification function 
(\ref{A(T)T>Tre}) is characterized by $\eta = \alpha_0 \frac{30}{16\pi^3 g_*}\frac{m^2 m_P^2}{T_F^4}$ and
$\nu = -4$. The value of $\eta \simeq 10^{-15}$ is required in order that results are consistent
with (\ref{alpha0bound}). The WIMPs mass still is $\simeq 10^2$GeV.



\begin{thebibliography}{99}

\bibitem{derham} C. de Rham, G. Gabadadze, Phys. Rev. D {\bf 82}, 044020 (2010). C. de Rham, G. Gabadadze, A.J. Tolley, Phys. Rev. Lett.
                    {\bf 106}, 231101 (2011).
\bibitem{pauli} M. Fierz and W. Pauli, Proc. Roy. Soc. Lond. a {\bf 173}, 211 (1939).
\bibitem{solartest} V. Zakharov, JETP Lett. {\bf 12}, 312 (1970). H. van Dam and M. Veltman, Nucl. Phys. B {\bf 22}, 397 (1970).
\bibitem{siegel} W. Siegel, Phys. Rev. D {\bf 49}, 4144 (1994). N. Arkani-Hamed, H. Georgi, and M.D. Schwartz, Ann. Phys. {\bf 305},
                    96 (2003).
\bibitem{deser} D.G. Boulware and S. Deser, Phys. Rev. D {\bf 6}, 3368 (1972).
\bibitem{volkov} A.H. Chamseddine, M.S. Volkov, Phys. Lett. B {\bf 704}, 652 (2011). M.S. Volkov, JHEP01 (2012) 035.
\bibitem{langlois} D. Langlois and A. Naruko, arXiv:1206.6810[hep-th].
\bibitem{gong} Y. Gong, arXiv:1207.1726[gr-qc].
\bibitem{strauss} M. von Strauss, A. Schmidt-May, J. Enander, E. Mortsell, and S.F. Hassan, arXiv:1111.1655[gr-qc].
\bibitem{kobayashi} T. Kobayashi {\it et al.}, arXiv:1205.4938[hep-th].
\bibitem{gratia} P. Gratia, W. Hu, and M. Wyman, arXiv:1205.4241[hep-th].
\bibitem{comelli} D. Comelli, M. Crisostomi, F. Nesti, L. Pilo, arXiv:1111.1983[hep-th].
\bibitem{damico} G. D'Amico {\it et al.}, Phys. Rev. D {\bf 84}, 124046 (2011).
\bibitem{lin} A.E. Gumrukcuoglu, C. Lin, and S. Mukohyama, JCAP {\bf 1111}, 030 (2011).
\bibitem{koyama} K. Koyama, G. Niz, and G. Tasinato, Phys. Rev. Lett. {\bf 107}, 131101 (2011); Phys. Rev. D {\bf 84}, 064033 (2011).
\bibitem{kurt} K. Hinterbichler, Rev. Mod. Phys. {\bf 84}, 671 (2012).
\bibitem{cardone} V. Cardone, N. Radicella, and L. Parisi, arXiv:1205.1613[astro-ph.CO].
\bibitem{PAMELA6} O. Adriani {\it et al.}, arXiv:0810.4995 [astro-ph].
\bibitem{kolb} E.W. Kolb, M.S. Turner, {\it The Early Universe}, Addison Wesley Publishing Company, 1989.
\bibitem{bernstein} J. Bernstein, L.S. Brown, G. Feinberg, Rev. Mod. Phys. {\bf 61}, 25 (1989).
\bibitem{torres} D.F. Torres, H. Vucetich, A. Plastino, Phys. Rev. Lett. {\bf 79}, 1588 (1997).
                 G. Lambiase, Phys. Rev. D {\bf 72}, 087702 (2005).
\bibitem{coc} A. Coc {\it et al.}, Astrophys. J. {\bf 600}, 544 (2004).
\bibitem{altriBBN} D. Kirkman {\it et al.}, Astrophys. J. Suppl. Ser. {\bf 149}, 1 (2003).
                Y.I. Izatov {\it et al.}, Astrophys. J. {\bf 527}, 757 (1999).
                Y.I. Izatov, T.X. Thuan, Astrophys. J. {\bf 602}, 200 (2004); Astrophys. J. {\bf 500}, 188 (1998).
                B.D. Fields, K.A. Olive, Astrophys. J. {\bf 506}, 177 (1998).
                K.A. Olive, E. Stillman, G. Steigman, Astrophys. J. {\bf 483}, 788 (1997).
\bibitem{fornengo} R. Catena, N. Fornengo, M. Pato, L. Pieri, A. Masiero, Phys. Rev. D  {\bf 81}, 123522 (2010).
                    M. Schelke, R. Catena, N. Fornengo, A. Masiero, M. Pietroni, Phys. Rev. D {\bf 74}, 083505 (2006).
\bibitem{BD1} R. Catena, N. Fornengo, A. Masiero, M. Pieroni, F. Rosati, Phys. Rev. D {\bf 70}, 063519 (2004).
            R. Catena, N. Fornengo, A. Masiero, M. Pietroni, M. Schelke, JHEP {\bf 10}, 003 (2008).
\bibitem{BD} M. Kamionkowski, M.S. Turner, Phys. Rev. D {\bf 42}, 3310 (1990).
            D.I. Santiago, D. Kalligas, R.V. Wagoner, Phys. Rev. D {\bf 58}, 124005 (1998).
            S. Profumo, P. Ullio, JCAP {\bf 0311}, 006 (2003).
            P. Salati, Phys. Lett. B {\bf 571}, 121 (2003).
            G. Gelmini, P. Gondolo, arXiv:1009.3690 [astro-ph.CO].
            S. Capozziello, M. De Laurentis, and G. Lambiase, Phys. Lett. B {\bf 715}, 1 (2012).
            G. D'Amico, M. Kamionkowski, K. Sigurdson, arXiv:0907.1912 [astro-ph.CO].
\end{thebibliography}
\end{document}